# Quantum key distribution system in standard telecommunications fiber using a short wavelength single-photon source

R. J. Collins,[a)][f)] P. J. Clarke,[a)] V. Fernández,[a)][d)] K. J. Gordon,[a)][e)]

M. N. Makhonin,[b)] J. A. Timpson,[b)] A. Tahraoui,[c)] M. Hopkinson,[c)] A. M. Fox,[b)]

M. S. Skolnick,[b)] and G. S. Buller[a)][g)]

[a)]School of Engineering and Physical Sciences, Heriot-Watt University, Riccarton, Edinburgh, EH14 4AS, United Kingdom

[b)]Department of Physics and Astronomy, University of Sheffield, Sheffield, S3 7RH, United Kingdom

[c)]Department of Electronic and Electrical Engineering, University of Sheffield, Sheffield, S1 3JD, United Kingdom

[d)]Present address: Instituto de Física Aplicada, Consejo Superior de Investigaciones Científicas (CSIC), Serrano 144, 28006 Madrid, Spain

[e)]Present address: SELEX Galileo, Ferry Road, Edinburgh, EH5 2XS, United Kingdom

[f)]Electronic mail: r.j.collins@hw.ac.uk

[g)]Electronic mail: g.s.buller@hw.ac.uk

A demonstration of the principles of quantum key distribution is performed using a single-photon source in a proof of concept test-bed over a distance of 2 km in standard telecommunications optical fiber. The single-photon source was an optically-pumped quantum dot in a microcavity emitting at a wavelength of 895 nm. Characterization of the quantum key distribution parameters was performed at a range of different optical excitation powers. An investigation of the effect of varying the optical excitation power of the quantum dot microcavity on the quantum bit error rate and cryptographic key exchange rate of the system are presented.

**I. INTRODUCTION**

Quantum key distribution (QKD) provides a verifiably secure means for two authorized parties (Alice and Bob) to share a cryptographic key.[1] The first complete QKD protocol was proposed by Bennett and Brassard in 1984[2] and progress since has been rapid, with key exchange having now being demonstrated at distances of up to 200 km in fiber[3] and distances of up to 144 km in free space.[4] The first gigahertz clock rate QKD demonstration in fiber[5] was published in 2004 and, to date, clock rates of up to 12 GHz have been demonstrated in fiber[3] and of up to 1.25 GHz in free space.[6] Technology has now progressed to the point where there are now several commercial systems available.[7] However, further research continues on different protocols, as well as on increasing the range, clock rate, security and stability under changing environmental conditions.

Single-photon sources offer potential security advantages over the commonly-used weak coherent pulses (WCP) - typically attenuated lasers - in QKD,[8] since WCP sources may have a multi-photon probability which can be sufficient to present a security risk in practical applications.[9]  If an eavesdropper (Eve) employs the photon number splitting (PNS) attack, she preferentially interrogates the multi-photon pulses to determine the quantum state.[9]  There has been much interest in the field of decoy states as a potential solution to this problem,[10] and this approach has been used in a number of experimental test-beds.[11-13]  However, unconditional security guaranteed by the fundamental laws of quantum physics still requires single-photon sources.[1, 9, 14]

Single-photon sources have been demonstrated in QKD using nitrogen vacancies in diamond (NVD) in free space transmission[15] and quantum dots in both free space[16] and optical fiber.[17] When compared to the clock frequencies now being demonstrated in WCP photon source QKD test-beds, many single-photon sources have been limited to low excitation pulse repetition rates.  However, recent work with electrically excited quantum dots has shown the promise of GHz excitation pulse repetition frequency operation.[18]  In addition, on-demand single-photon emission rates as high as 31 MHz have been demonstrated from quantum dot microcavities.[19]

The test-bed presented in this paper operates at a wavelength of 895 nm, as opposed to the telecommunications wavelengths at 1.3 μm and 1.55 μm.  Short wavelengths have the advantage that they are spectrally separated from the dense classical telecommunication traffic already present in telecommunications networks.[20] In addition, the use of the lower wavelength permits the use of silicon single-photon avalanche diodes (Si-SPADs), with their lower dark count rates and afterpulsing probabilities compared to the longer wavelength alternatives.  However, the losses incurred in standard telecommunications optical fiber at these wavelengths are significantly higher than those incurred for longer wavelengths, typically several dBkm$^{-1}$ as opposed to ~0.2 dBkm$^{-1}$ at a wavelength of 1550 nm. This increased loss limits transmission distances to those consistent with metropolitan telecommunications access network links, i.e. typically less than 20 km.[20, 21]

In this paper we demonstrate a single-photon source based on a quantum dot microcavity which has been applied to a polarization-based BB84 protocol quantum key distribution test-bed in standard telecommunications optical fiber. The test-bed is demonstrated with Alice and Bob directly connected to each other and also with 2 km of standard telecommunications optical fiber serving as a quantum transmission channel. This approach examines the potential of relatively short wavelength (i.e. $\lambda < 1000$ nm) quantum dot sources in optical fiber-based quantum key distribution.

## II. SHORT WAVELENGTH SINGLE-PHOTON SOURCE

Quantum dot microcavities were grown with a cavity consisting of 16 pairs of distributed Bragg reflector (DBR) mirrors below an InAs dot layer and 6 above.[22] These microcavities emitted photons in the wavelength range 890 nm to 905 nm. A custom-designed microscope was developed in-house to examine the quantum dot microcavities and couple the emission into an optical fiber,[23] as shown in Fig. 1. The quantum dots were optically excited by a ~ 90 ps FWHM pulse duration, 784 nm wavelength semiconductor diode laser. For the results presented in this paper, the pulse repetition frequency was 40 MHz. The cryostat was maintained at a temperature of 53 K to observe optimal single-photon emission from the quantum dots. A laser with an identical emission wavelength to the transitions in the quantum dot was transmitted through the microscope to evaluate its coupling efficiency. Using this technique, it was calculated that the microscope was able to couple ~11% of the photons emitted by the microcavity into the acceptance cone angle of the 0.42 NA microscope objective into the 9 μm core diameter exit optical fiber.

A quantum dot within a 1 μm diameter cylindrical pillar (containing approximately 10 to 100 quantum dots) was selected as the best candidate for use in short wavelength single-photon quantum key distribution, based on spectral emission, decay time and second order autocorrelation ($g^{(2)}(0)$). Fig. 2 shows a photoluminescence spectrum from the micropillar, obtained at an excitation power of 1 μW (measured at the cryostat window). The spectra presented in Fig. 2 were measured using a liquid nitrogen cooled front illuminated charge coupled device (CCD) connected to a 0.5 m imaging triple grating monochromator. The CCD had a dark count rate of ~750 counts per pixel per second, leading to the background level of ~0.4 evident in the figure. The dark gray plots in the main Fig. 2 and insert clearly show the multiple spectral lines corresponding to many transitions and many dots. The light gray plot in the insert shows the same spectrum after spectral filtering using two narrow-bandpass filters to isolate one single transition in a single quantum dot. This transition was selected because it gave usable $g^{(2)}(0)$ values over a range of excitation powers from 0.25 μW to 5 μW, as shown in Fig. 3.

The $g^{(2)}(0)$ values were measured using a Hanbury Brown and Twiss (HBT) experiment in optical fiber.[24] The $g^{(2)}(0)$ value varies with excitation power: at an excitation power of 0.25 μW the $g^{(2)}(0)$ was 0.32 and at an excitation power of 5 μW the $g^{(2)}(0)$ increased to 0.85. The detection efficiency of the commercially available thick junction Si-SPADs[25] used to detect the photons was measured to be 40% at the emission wavelength of the quantum dot. From this value and the coupling efficiency of the microscope it was possible to calculate that the photon emission rate contained within the acceptance angle cone of the microscope objective was 200 kHz for a $g^{(2)}(0)$ of 0.32 rising to 4 MHz at a $g^{(2)}(0)$ of 0.85.

The dot emission lifetimes were characterized using time-resolved photoluminescence (TRPL) techniques[23]. Fig. 4 shows a dot decay trace at an excitation power of 1 µW, corresponding to a $g^{(2)}(0)$ of 0.39, with the instrumental response shown for comparison. The photon emission rate from the quantum dot microcavity (after correction for coupling loss in the microscope and detection efficiency) at this excitation power was 480 kHz. An iterative reconvolution technique[26] was used to measure the primary photoluminescence (PL) lifetime of this emission as 464 ps. The insert shows the same dot decay over a longer timescale. The evident long tail with a decay time (of >300 ns) is believed to be caused by spin flip and consequent formation of dark states, which then reappear after a second spin flip.[27].The longest PL lifetime observed from this microcavity was 563 ps at an excitation power of 5 µW.

## III. SINGLE-PHOTON QUANTUM KEY DISTRIBUTION

The experimental system used for single-photon quantum key distribution is shown schematically in Fig. 5. The photons from the quantum dot microcavity were focused through a free-space resonant frequency polarization modulator which was used to set the quantum states of the BB84 protocol in polarization at a clock frequency of 40 MHz. A high extinction ratio (10,000:1) fixed polarizer was used to ensure that the randomly polarized photons emitted by the quantum dot were highly linearly polarized prior to their polarization state being modulated to each of the desired states by the modulator. A resonant frequency, free-space polarization modulator[28] based on a crystal of MgO:LiNbO$_3$ was selected for this application as it was capable of modulating photons of the emission wavelength at the desired clock frequency with a V$_\pi$ voltage of 19V. The modulator required a focused beam with diameter less than 500 μm along the 56 mm length of the crystal to ensure efficient polarization modulation and this was achieved by using a 16 mm focal length lens to collimate the photons emitted from the 9 μm core diameter fiber and a 500 mm focal length lens to focus the photons through the modulator. The modulator was driven using a commercially available electrical amplifier[29] designed to operate at 40 MHz and able to set the output driving voltage (corresponding to polarization state) depending on the analogue voltage level at a 50 Ω terminated input. Transmission of the photons through the free-space polarization modulator and associated optics incurred total losses of 10.6 dB. When Alice and Bob were directly connected, the ratio of the major to minor axis of the linear states in the BB84 protocol was measured to be 545:1 at the polarizing beam splitters in Bob.

The quantum channel was composed of 9 μm diameter core standard telecommunications fiber (Corning SMF-28e®) which was single-mode at wavelengths in the range ~1250 nm to ~1625 nm.[30] It is possible to suppress the propagation of the higher order modes which will propagate in the fiber when it is illuminated with photons of wavelength 895 nm by splicing short (<1 m) length of 5 μm core diameter fiber onto the standard telecommunications fiber.[31] The gray crosses in Fig. 5 denote the points at which 5 μm core diameter fiber was fusion spliced onto the 9 μm core diameter fiber. At the wavelength of emission of the quantum dot, the standard telecommunication optical fiber quantum channel exhibited losses of ~2.2 dBkm$^{-1}$.

At Bob, a 50/50 beamsplitter acts as a random routing component performing the basis set selection of the BB84 protocol. Static polarization controllers operating through mechanically induced birefringence served to align the polarization states with the transmission or reflection axis of polarization dependent beamsplitters (PBS). Further to the protocol loss of 3 dB incurred during basis set reconciliation, there was an additional 4.76 dB of loss introduced by imperfections in Bob's optical components.

A figure of merit for a QKD system is the quantum bit error rate (QBER)[32]. In our test-bed Bob uses free-running detectors and then gates the events in software using a gate of 300 ps duration around the expected bit times to reduce the effects of dark counts on the QBER. The QBER can be calculated from:

$$Q = \frac{N_I}{N_C + N_I} \qquad (1)$$

where $Q$ is the QBER. $N_C$ is the number of detector events which occur within all of the 300 ps gates across all of the detectors and are caused by the correct quantum state. $N_I$ is the number of detector events which occur within the same time windows but are not due to the correct quantum state. $N_I$ contains components from both detector dark noise and polarization leakage. The long decay tail evident on the quantum dot emission shown in Fig. 4 decreases the number of photons emitted from the source which are contained within the 300 ps duration time window. This leads to an increase in the significance of the dark count rate of the detectors (300 Hz) in the calculation of QBER.

Measurements were made of QBER at a range of different excitation powers, and the results can be seen for transmission distances of 0 km and 2 km in Fig. 6. It can be seen from Fig. 6 that at higher excitation powers, corresponding to increased photon fluxes, the QBER decreases. The lowest QBER measured using a $g^{(2)}(0) < 1$ was 1.22% at 0 km and 6.21% at 2 km for a $g^{(2)}(0)$ of 0.85. At the lowest $g^{(2)}(0)$ of 0.32 for 0 km the QBER was 21.9%.

A simple analysis of the final key exchange rate may be made by examining the effect of the CASCADE error correction protocol[33] on the measured click rate of the detectors. The effect of the CASCADE error correction protocol on the sifted bit rate can be calculated from:

$$R_{net} = \left(1 - f_p H_2(Q)\right) R_{Sifted} \qquad (2)$$

where $Q$ is the quantum bit error rate defined earlier, $R_{net}$ is the net bit rate, $R_{sifted}$ is the sifted bit rate after temporal filtering, $f_p$ is a measure of the additional inefficiency of the error correction protocol when compared to the theoretical Shannon limit and $H_2(Q)$ is the binary entropy function given by:[34]

$$H_2(Q) = -Q \log_2(Q) - (1 - Q) \log_2(1 - Q). \qquad (3)$$

For the CASCADE error correction protocol it can be determined[35] that $f_p = 1.16$. The filled points in Fig. 7 shows the calculated net bit-rate against the excitation power for both transmission distances when only considering the CASCADE error correction protocol.

This simple analysis does not take into account the additional bits which must be sacrificed during secure key distillation to compensate for the PNS attack. A more complete analysis based on the GLLP technique for imperfect devices leads to the following definition for secure net bit rate:[36]

$$R_{net} = \left( (1 - \Delta) - f_p H_2(Q) - (1 - \Delta) f_p H_2\left( \frac{Q}{1 - \Delta} \right) \right) R_{Sifted} \qquad (4)$$

where $\Delta$ is the fraction of the bits transmitted by Alice which are intercepted by the eavesdropper. To establish a lower boundary on the secure key exchange rate, it is necessary to assume that the eavesdropper intercepts every multi-photon pulse emitted by Alice[17] so that $\Delta \approx g^{(2)}(0)\frac{\mu^2}{2}$. The results of the application of the GLLP based are shown by the unfilled points in Fig. 7. It can be seen from Fig. 7 that there are a number of conditions for both transmission distances which are secure against the PNS attack when analyzed using the GLLP technique for imperfect devices.

## IV. DISCUSSION AND CONCLUSIONS

A proof-of-concept quantum key distribution system using photons emitted at a wavelength of 895 nm by a quantum dot microcavity has been demonstrated using quantum channels composed of up to 2 km of standard telecommunications optical fiber for a range of different excitation powers. The highest photon emission rate from the quantum dot microcavity was calculated to be 4 MHz at an excitation power of 5 μW and a $g^{(2)}(0)$ of 0.85, once the 11% coupling efficiency of the microscope and the 40% detection efficiency of the Si-SPADs were taken into account. The maximum value observed for the primary excited state lifetime of the dot (563 ps at an excitation power of 5 μW) sets a limit on the maximum excitation pulse repetition rate which is in the region of several hundred MHz. However, the polarization modulator used in these experiments was incapable of operating at such frequencies due to the relatively large volume of the MgO:LiNbO$_3$ crystal. The highest emission rate reported for a quantum dot microcavity was 31 MHz (after correction for detection efficiency) at a $g^{(2)}(0)$ of 0.4 and an excitation pulse repetition rate of 82 MHz by Strauf et al..[19]  This microcavity emitted photons at a wavelength of 916 nm which is compatible with the apparatus used for the experiments detailed in this paper.  If such a single-photon source was integrated into our system using a modulator capable of giving similar performance at a clock rate of 82 MHz, simulations with a WCP based source at the same photon flux indicate that at a transmission distance of 2 km the observed QBER would be reduced to be 1.08%.  Calculations using the GLLP technique[36] and a $g^{(2)}(0)$ of 0.4 predict that the secure key exchange rate would be 288 bits[-1].   Replacing the free-space polarization modulator with an in-line fiber coupled version could significantly reduce the loss incurred due to coupling light from a fiber. If the free-space polarization modulator was replaced with an optimized in-line fiber module, the loss at this component could be reduced to ~3 dB.   In this case, with the microcavities used by Strauf et al.,[19] simulations using a WCP based source at the same photon flux indicate that the QBER at 2 km would be 0.68% and the secure bit-rate would increase to 1712 bits[-1] for a $g^{(2)}(0)$ of 0.4.  These bit rates offer the prospect of using, for example, the RSA encryption algorithm[37] with a 1024 bit key which is refreshed every second.  It is not necessary to utilize the key as it is generated and it may be stored until sufficient secure bits have been acquired for future use.

Examining the results from the quantum dot microcavities presented in this paper, it is possible to calculate a maximum transmission distance if the microcavities of Strauf et al. were integrated into the revised QKD test-bed.  The QBER rises above[38] 11% for photon fluxes less than that achieved for an excitation power of 1 µW at a distance of 2 km.  By calculating the loss of the quantum channel required to reduce the photon flux at Bob to a level comparable to that observed with our microcavities at an excitation power of 1μW and a distance of 2 km, we have estimated the maximum transmission distance achievable with the revised test-bed is 11 km using the microcavities of Strauf et al.. In all cases, this key exchange works independently of any data channels operating simultaneously at longer wavelengths on these fiber links.

This test-bed has demonstrated the potential of a short wavelength single-photon source in quantum key distribution over standard telecommunications fiber. It can be seen that with improvements in the photon emission rate at low $g^{(2)}(0)$ values, this approach may offer the prospect of higher secure bit-rate transmission over longer lengths of standard telecommunications optical fiber.

**ACKNOWLEDGEMENTS**

The authors acknowledge the support of UK Engineering and Physical Sciences Research Council project GR/T09392 and GR/T09408. R. J. Collins, P. J. Clarke and G. S. Buller are part of the Scottish Universities Physics Alliance (SUPA).

**FIGURE CAPTIONS**

FIG. 1. The microscope used to image the quantum dot microcavity samples. The cryostat is fixed below a movable baseplate containing the imaging optics. The white light source and camera are used to image the sample prior to measurements. Optical excitation of the sample is provided by the 784 nm wavelength laser diode which is reflected by a BK7 plate glass beamsplitter and a gold-faced mirror to the sample via a ×50 microscope objective with a numerical aperture of 0.42. The photons emitted by the dot in the microcavity are collected into a 9 μm core diameter optical fiber which can be connected to additional characterization experiments (e.g. for $g^{(2)}(0)$ measurements) or to the polarization modulator for QKD. This system coupled ~11% of the photons emitted from the microcavity into the acceptance cone angle of the microscope objective into the 9 μm core diameter fiber.

FIG. 2. A normalized spectrum obtained from the quantum dot microcavity single-photon source at an excitation power which gives a $g^{(2)}(0)$ of 0.38, corresponding to a photon emission rate from the quantum dot microcavity (after correction for detection efficiency) of 480 kHz. The insert shows a subset of the same spectrum before spectral filtering (dark gray) and after (light gray). These measurements were performed using a liquid nitrogen cooled front illuminated charge coupled device (CCD) connected to a 0.5 m imaging triple grating monochromator. The background level of ~0.4 is due to the dark noise level of the CCD which was approximately 750 counts per pixel per second, significantly higher than the 300 counts per second dark count rate of the Si-SPADs used to measure the autocorrelation value and QKD bit-rates.

FIG. 3.  The photon flux exiting Alice against the excitation power measured at the cryostat window.  The right-hand axis indicates the photon flux emitted from the microcavity into the acceptance cone angle of the microscope objective, assuming 40% detection efficiency for the silicon single-photon avalanche diodes and an ~11% coupling efficiency in the microscope.  The values specified next to the data points denote the value of the second order autocorrelation function ($g^{(2)}(0)$) at these excitation powers.

FIG. 4.  A normalized time-resolved photoluminescence (TRPL) trace of the quantum dot at an excitation power which gives a $g^{(2)}(0)$ of 0.39 (gray line) and a normalized instrumental response (dark gray line).  The photon emission rate of the microcavity, after correction for detection efficiency, was 480 kHz.  An iterative reconvolution technique[26] was used to measure the primary photoluminescence lifetime of this emission as 464 ps.  The insert shows the same quantum dot TRPL result over a long timescale, clearly showing the long decay tail.

FIG. 5. A schematic diagram of the short wavelength quantum key distribution test-bed.  The box labeled "Single-Photon Source" contains the custom microscope.  SPC is a static polarization controller, Si-SPAD is a silicon avalanche photodiode and the gray crosses denote the points at which the 5 μm core diameter fiber in Alice and Bob is spliced to the 9 μm core diameter standard telecommunications fiber which comprises the quantum channel.

FIG. 6.   The quantum bit error rate (QBER), expressed as a percentage, against excitation power measured at the cryostat window for the quantum dot microcavity single-photon source in a BB84 quantum key distribution system.  The black triangles (▲) denote a transmission distance of 0 km while the gray circles (●) denote a transmission distance of 2 km.  The values specified next to the data points denote the value of the second order autocorrelation function ($g^{(2)}(0)$) at these excitation powers.

FIG. 7.   The filled points denote the net bit-rate against excitation power for the quantum dot microcavity single-photon source in a BB84 quantum key distribution system at distances of 0 km and 2 km analyzed using the Cascade error correction protocol[33].  The black triangles (▲) denote a transmission distance of 0 km while the gray circles (●) denote a transmission distance of 2 km.  The unfilled points denote the transmission conditions which were found to be secure against the PNS attack[9] when analyzed using the GLLP[36] technique with triangles (Δ) denoting a transmission distance of 0 km and circles (○) a transmission distance of 2 km.

**Fig. 1 – One column**

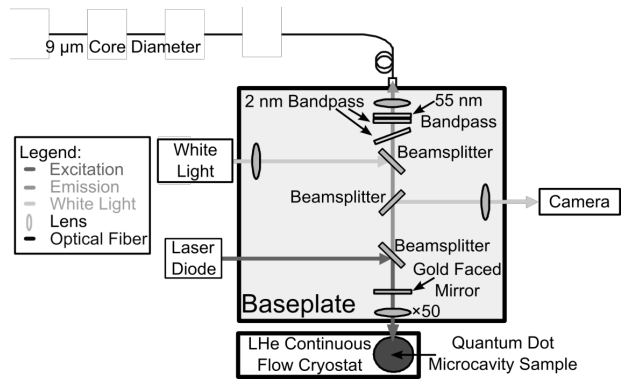

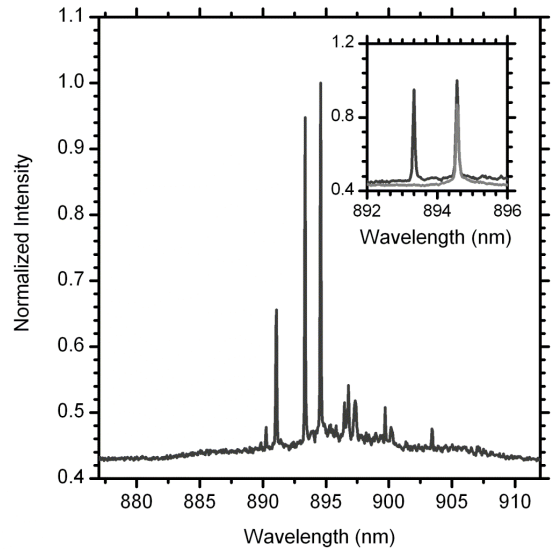

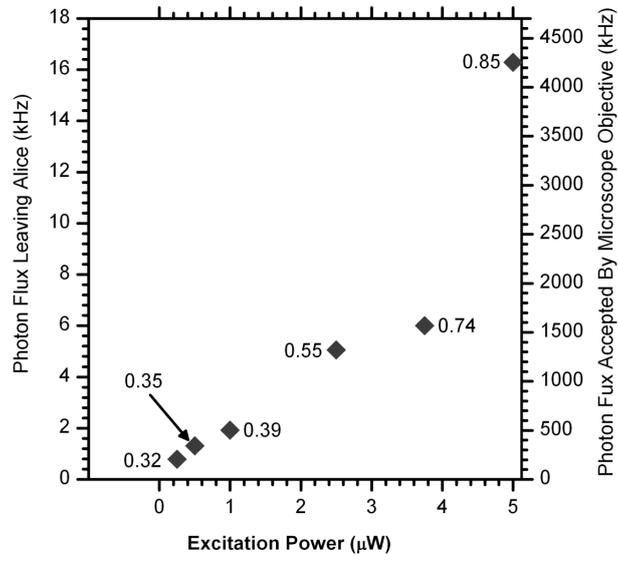

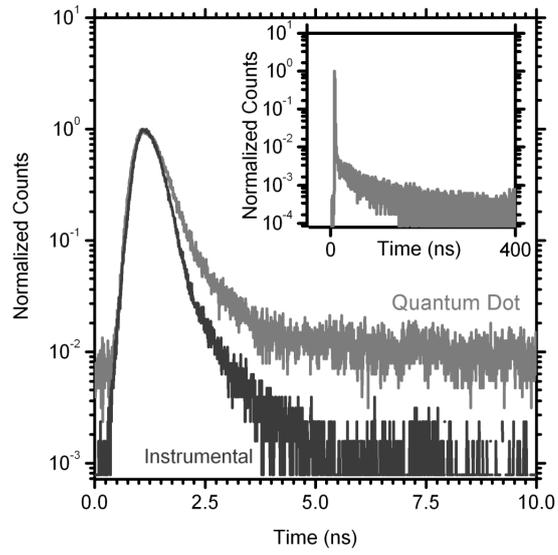

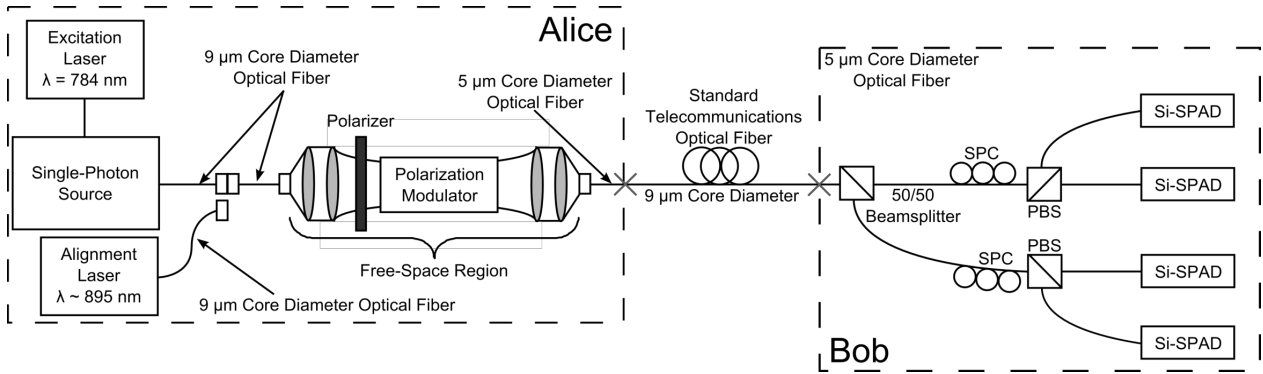

**Fig. 6 – One column**

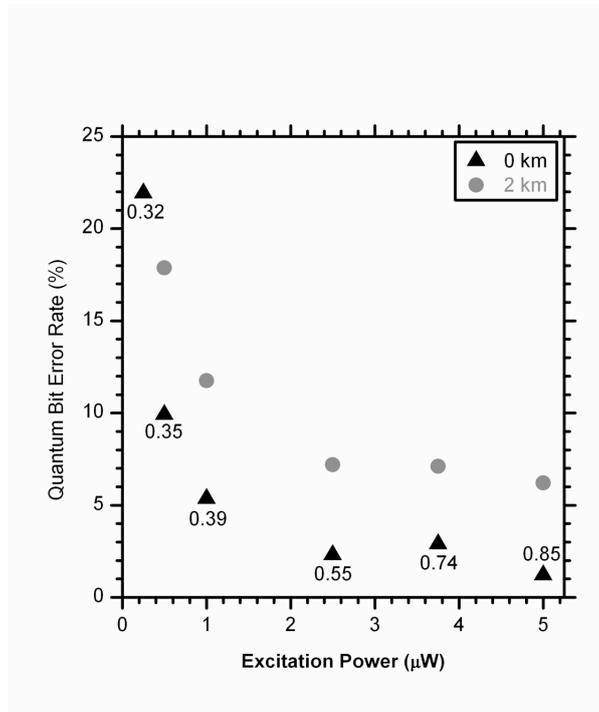

**Fig. 7 - One column**

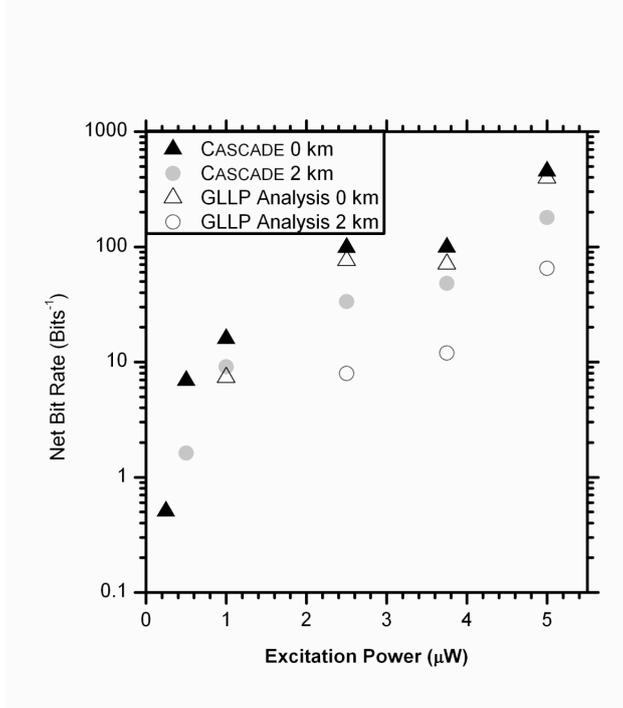